\documentclass[a4paper,11pt]{article}
\usepackage{pos}
\usepackage{tikz}
\usepackage[compat=1.1.0]{tikz-feynman}
\usetikzlibrary{decorations.markings}

\title{Inclusive semileptonic $B$-decays from lattice QCD}
\author[a,b]{Paolo Gambino}
\author[c,d]{Shoji Hashimoto}
\author[a,e]{Sandro M\"achler}
\author[a]{Marco Panero}
\author[f]{Francesco Sanfilippo}
\author[f]{Silvano Simula}
\author*[a]{Antonio Smecca}
\author[g]{Nazario Tantalo}

\affiliation[a]{Dipartimento di Fisica, Universit\`a di Torino \& INFN, Sezione di Torino,\\Via Pietro Giuria 1, I-10125 Turin, Italy}
\affiliation[b]{Max Planck Institute for Physics, F\"ohringer Ring 6, 80805 M\"unchen, Germany}
\affiliation[c]{Theory Center, Institute of Particle and Nuclear Studies, High Energy Accelerator Research Organization (KEK), Tsukuba 305-0801, Japan}
\affiliation[d]{School of High Energy Accelerator Science, The Graduate University for Advanced Studies (SOKENDAI), Tsukuba 305-0801, Japan}
\affiliation[e]{Physikinstitut, Universit\"at Z\"urich, Winterthurerstrasse 190, CH-8057 Z\"urich, Switzerland}
\affiliation[f]{INFN, Sezione di Roma Tre, Via della Vasca Navale 84, I-00146 Rome, Italy}
\affiliation[g]{Dipartimento di Fisica, Universit\`a di Roma ``Tor Vergata'' \& INFN, Sezione di Roma ``Tor Vergata'', Via della Ricerca Scientifica 1, I-00133 Rome, Italy}

\emailAdd{antonio.smecca@unito.it}

\abstract{We present the lattice QCD calculation of inclusive semileptonic $B_s$-meson decays. We follow a recently proposed method, which is based on the extraction of smeared spectral densities from Euclidean correlation functions and on the numerical reconstruction of the integration kernel relevant for the inclusive decay rate calculation. We compute four-point Euclidean correlation functions using JLQCD and ETM gauge ensembles with unphysically light $b$-quark masses, and apply two different methods for the integration kernel reconstruction. Finally, we show that the lattice results obtained in this work are in good agreement with the analytic predictions of the operator-product-expansion. This opens the path for a future full lattice QCD calculation to be used as theoretical input for the determination of the magnitude of the CKM element $V_{cb}$.}

\FullConference{%
The 39th International Symposium on Lattice Field Theory,\\
8th-13th August, 2022,\\
Rheinische Friedrich-Wilhelms-Universität Bonn, Bonn, Germany
}

\begin{document}
\maketitle

\section{Introduction}

The study of the semileptonic decays of heavy mesons is an important and active area of research since their decay rates are proportional to the modulus squared of the elements of the Cabibbo-Kobayashi-Maskawa (CKM) mixing matrix~\cite{ParticleDataGroup:2020ssz, HFLAV:2019otj}. 
%proportional to the modulus of CKM. The determination of CKM is important because...
The CKM matrix is a complex unitary matrix in the Standard Model (SM) and the precise determination of its elements is very important in order to validate the SM, as any deviation from unitarity would suggest the effect of new physics.
For this reason, one of the main goals of flavour physics is to overconstrain the CKM elements in order to check for any such deviations.
At the moment, one of the most persistent tensions is the $\simeq 3 \sigma$ discrepancy between exclusive and inclusive determinations of the magnitude of $V_{cb}$~\cite{Gambino:2019sif,Gambino:2020jvv}.
The determination of CKM elements is not easy as they cannot be simply calculated from theory, instead, one has to combine experimental measurements with theoretical non-perturbative calculations.

Lattice QCD has been extremely successful in providing precise and systematically improvable calculations for the form factors of exclusive semileptonic decays~\cite{FlavourLatticeAveragingGroup:2019iem}, while computations of inclusive quantities relied on the analytic method known as the Operator Product Expansion (OPE)~\cite{Wilson:1969zs, Kadanoff:1969zz}.
In this contribution we present our latest work \cite{Gambino:2022dvu}, regarding one of the first studies of inclusive semileptonic decays of heavy mesons using lattice QCD. This work is based on recently proposed methods \cite{Hashimoto:2017wqo,Gambino:2020crt} which allow for a complete first-principle computation of inclusive semileptonic decay rates starting from Euclidean correlation functions obtained from lattice simulations.
Although it is still a preliminary exploratory study, our results obtained from the lattice show a good agreement with the analytic predictions of the OPE, making us optimistic for a future full lattice study.
A complete lattice computation of inclusive decays would be an important step towards a better understanding of the $V_{cb}$ problem.
%The study of heavy mesons semileptonic decays is an active area of research since they encode direct information on the modulus of the elements of the Cabibbo-Kobayashi-Maskawa (CKM) mixing matrix. The CKM elements are fundamental parameters of the Standard Model (SM)

\section{Theoretical framework}

We follow the formalism outlined in ref.~\cite{Gambino:2020crt}, in which we consider the semileptonic decay of a $B_s$ meson into a pair of massless leptons $l\bar{\nu}$ and to charmed final states $X_c$. Working in the $B_s$ meson rest frame, we define the differential decay rate as
\begin{align}
  \frac{d\Gamma}{dq^2dq^0dE_\ell}=
  \frac{G_F^2|V_{cb}|^2}{8\pi^3} L_{\mu\nu}W^{\mu\nu},
  \label{eq:differential}
\end{align}
where $L_{\mu\nu}$ is the leptonic tensor and $W_{\mu\nu}$ is the hadronic tensor. The latter can be written in its spectral representation
\begin{align}
  W_{\mu\nu}(\omega,\boldsymbol{q}) = \frac{(2\pi)^3}{2m_{B}}\langle\bar{B}(\boldsymbol{0})|
  J_\mu^\dagger(0) \delta(\hat{H}-\omega)\delta^3(\hat{\boldsymbol{P}}+\boldsymbol{q})
  J_\nu(0)|\bar{B}(\boldsymbol{0})\rangle,
  \label{eq:spectre}
\end{align}
where $\hat{H}$ is the QCD Hamiltonian and $\hat{\boldsymbol{P}}$ is the momentum operator, while $J_{\mu}$ is the flavour changing current.

The hadronic tensor can be decomposed into Lorentz-invariant structure functions $Z^{(l)}(\omega,\boldsymbol{q}^2)$, such that after integrating over the lepton energy $E_l$ the differential decay rate can be rewritten as
\begin{align}
  \frac{d\Gamma}{dq^2}=\frac{G_F^2|V_{cb}|^2}{24\pi^3|\boldsymbol{q}|}\sum_{l=0}^2 \Big(\sqrt{q^2}\Big)^{2-l}Z^{(l)}(\boldsymbol{q}^2),
  \label{eq:IncDecay}
\end{align}
where $\omega$ is the energy of the final meson state in the rest frame of the $B_s$ meson and $Z^{(l)}(\boldsymbol{q}^2)$ is the integral over phase space of the structure functions $Z^{(l)}(\omega,\boldsymbol{q}^2)$ regulated by the integration kernel $\Theta^{l}(\omega_{max}-\omega)$:
\begin{align}
  Z^{(l)}(\boldsymbol{q}^2) = \int_0^\infty d\omega  \Theta^l(\omega_{\max}-\omega)Z^{(l)}(\omega,\boldsymbol{q}^2).
  \label{eq:Zfunction}
\end{align}
Equation~(\ref{eq:Zfunction}) plays a central role in our effort to calculate the differential decay rate showed in eq.~(\ref{eq:IncDecay}). For this reason, this work focuses on calculating this object using Euclidean correlators computed in lattice QCD simulations.

\section{Lattice computation}

The first step towards a lattice calculation of the inclusive semileptonic decay rate is to make a connection between the hadronic tensor in eq.~(\ref{eq:spectre}) and an appropriate Euclidean correlation function.
Indeed, while two- and three-point correlation functions are sufficient for accessing the ground state of a particle, required for exclusive calculations, they cannot access the full spectrum of charmed final states, which can instead be achieved by four-point correlation functions as showed in ref.~\cite{Gambino:2020crt}.
The four-point correlator can be written explicitly as
\begin{align}
  C_{\mu\nu}(t_{\mathrm{snk}},t_2,t_1,t_{\mathrm{src}};\boldsymbol{q}) =        
  \int d^3x\, e^{i\boldsymbol{q}\cdot\boldsymbol{x}}\, 
  T\langle 0\vert\, \tilde{\phi}_{B_s}(\boldsymbol{0};t_{\mathrm{snk}})
  J_\mu^\dagger(\boldsymbol{x};t_2) J_\nu(\boldsymbol{0};t_1)
  \tilde{\phi}_{B_s}^\dagger(\boldsymbol{0};t_{\mathrm{src}})\, \vert 0\rangle ,
  \label{eq:4PtCorrelator}
\end{align}
where the two currents are sandwiched between the $B_s$ meson states.

\begin{figure}
  \centering
  \begin{tikzpicture}
    [
      roundnode/.style={circle, draw=blue!60, very thick, minimum size=1.2cm},%fill=green!5
      squarednode/.style={rectangle, draw=blue!60, very thick, minimum size=8mm},
      decoration={
        markings,
    mark=at position 0.5 with {\arrow{stealth}}}
    ]
    %Nodes
    \node[roundnode, label={[xshift=-0.5cm,yshift=-1.5cm]$t_{snk}$}]      (Bmeson)                  {$B_s$};
    \node[squarednode, label={[yshift=-1.5cm]$t_2$}] (current_d) [above right =1cm and 1.3cm of Bmeson] {$J_{\mu}^{\dagger}$};
    \node[squarednode, label={[yshift=-1.5cm]$t_1$}]  (current)   [above right =1cm and 4.3cm of Bmeson] {$J_{\nu}$};      
    \node[roundnode, label={[xshift=0.7cm,yshift=-1.5cm]$t_{src}$}] (Dmeson)       [right=6cm of Bmeson] {$B_s$};

    %Lines
    \draw[very thick,postaction={decorate}] (current_d.west) .. controls +(left:1.5cm) and +(up:3mm)  .. (Bmeson.north) node[midway,above]  {$b$};
    \draw[very thick,postaction={decorate}] (Bmeson.south) .. controls +(down:1.3cm) and +(down:1.3cm)  .. (Dmeson.south) node[midway,above] {$\bar{s}$};
    \draw[very thick,postaction={decorate}] (Dmeson.north) .. controls +(up:3mm) and +(right:1.5cm)  .. (current.east) node[midway,above] {$b$};
    \draw[very thick,postaction={decorate}] (current.west) -- (current_d.east) node[midway,above] {$c$};

  \end{tikzpicture}
  \caption{Schematic representation of the four-point Euclidean correlation function defined in eq.~(\ref{eq:4PtCorrelator})}
\end{figure}
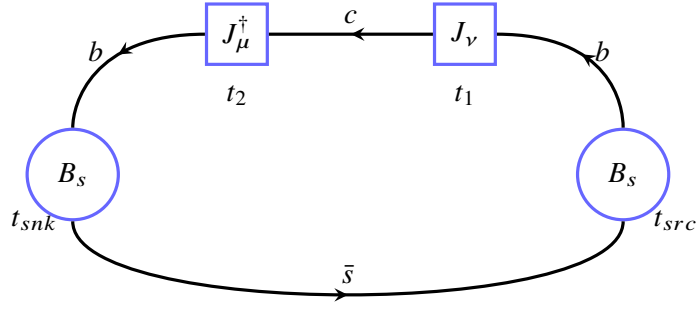
The Euclidean correlator is saturated by the matrix element
\begin{align}
  M_{\mu\nu}(t;\boldsymbol{q}) = e^{-m_B t}\, \int d^3x\, 
  \frac{e^{i\boldsymbol{q}\cdot\boldsymbol{x}}}{2m_B}
  \langle \bar B(\boldsymbol{0})|J_\mu^\dagger(\boldsymbol{x},\!t)J_\nu(\boldsymbol{0},\!0)
  |\bar B(\boldsymbol{0})\rangle,
  \label{eq:mat_element}
\end{align}
which can be shown, after a few lines of algebra, to be equal to the integral over the hadronic tensor,
\begin{align}
  M_{\mu\nu}(t;\boldsymbol{q})=\int_0^{\infty}d\omega W_{\mu\nu}(\omega,\boldsymbol{q})\,e^{-\omega t}.
\end{align}

In our analysis we focus on the contribution to the correlator between $t_1$ and $t_2$, in which the multi-particle states of the spectrum propagate. We also write a linear combination of the Euclidean correlator in terms of the hadronic-tensor components $Z^{(l)}(\omega,\boldsymbol{q})$ as
\begin{align}
  G^{(l)}(a\tau;\boldsymbol{q}) = \int_0^{\infty} d\omega Z^{(l)}(\omega,\boldsymbol{q}^2)e^{-\omega a\tau} ,
  \label{eq:Inverse_problem}
\end{align}
where we expressed the Euclidean time in units of the lattice spacing (denoted by $a$) as $a\tau$.

In our work we used two gauge ensembles generated by the JLQCD collaboration~\cite{Nakayama:2017lav, Colquhoun:2022atw} and by the ETM collaboration~\cite{Baron:2010bv, ETM:2010cqp, Frezzotti:2000nk, Frezzotti:2003xj, Frezzotti:2003ni, EuropeanTwistedMass:2014osg}, which use Domain Wall fermions and Twisted-Mass Wilson fermions respectively. In the lattice simulations neither collaboration is able to accommodate a relativistic $b$ quark with its physical mass. As a consequence, our results are obtained with an unphysically light $B_s$ meson.
\begin{table}
  \begin{tabular}{c c c c c }
    \hline
    $L^3 \times T$ & $N_{cnfg}$ & $a$ $(fm)$& &\\
    $32^3 \times 64$ & $150$ & $0.0815(30)$& &\\
    \hline
    $a\mu_{l} $ & $a\mu_{\sigma} $ & $a\mu_{\delta} $ & $m_{\pi}$ $(\text{\rm MeV})$ &\\
    $0.0025$ & $0.135$ & $0.170$  & $375(13)$ \\
    %      \hline
    %      $a\mu_s $ & $a\mu_c$ & $a\mu_b$ & &\\
%      $0.021$ & $0.25$ & $0.5$ & &\\
    \hline
  \end{tabular}
  \quad
  \begin{tabular}{c c c c c }
    \hline
    $L^3 \times T \times L_s$ & $N_{cnfg}$ & $a$ $(fm)$ & &\\
    $48^3 \times 96 \times 8$ & $100$ & $0.055$& &\\
    \hline
    $am_{ud}$ & $am_{s}$ & $m_{\pi} (\text{\rm MeV})$ &\\
    $0.0042$ & $0.0025$ & $300(1)$ & \\    
    \hline    
  \end{tabular}
  \caption{Gauge ensemble details for the ETM (left) and JLQCD (right) collaborations. The number of dynamical quarks is $N_f=2+1+1$ in ETMC case and $N_f=2+1$ in JLQCD case. In the Twisted Mass simulation, $a\mu_{\sigma}$ and $a\mu_{\delta}$ are heavy doublet parameters related to the strange- and charm-quark masses.}
  \label{tab:lattice}  
\end{table}
In table~\ref{tab:lattice} the details of the gauge configurations are summarised. Further details about the ensembles can be found in ref.~\cite{Gambino:2022dvu} and references therein.

\section{Kernel reconstruction}\label{sec:kernel}

In order to compute the differential decay rate one has to extract the hadronic spectral density from Euclidean correlators, which is an ill-posed inverse problem and therefore needs to be tackled with appropriate numerical methods.
In this work we apply two methods, one based on Chebyshev polynomials~\cite{Bailas:2020qmv} and another one based on the variant of the Backus-Gilbert method proposed in ref.~\cite{Hansen:2019idp}, which are applied respectively to the JLQCD ensemble and the ETMC ensemble.
The first step in both methods is to use the fact that any smooth function $f(\omega)$ can be approximated numerically by a series of polynomials
\begin{align}
  f(\omega) = \sum_{\tau}^{\infty}g_{\tau}e^{-a\omega\tau},
\end{align}
so that eq.~(\ref{eq:Inverse_problem}) can be rewritten as
\begin{align}
  \sum_{\tau}^{\infty}g_{\tau}G^{(l)}(a\tau;\boldsymbol{q}) = \int_0^{\infty} d\omega Z^{(l)}(\omega;\boldsymbol{q}^2)f(\omega)\;.
\end{align}
One can then notice how the term on the r.h.s. of the above equation is analogous to the expression in eq.~(\ref{eq:Zfunction}), if the kernel $\Theta^l(\omega_{max}-\omega)$ is substituted with $f(\omega)$.
However, since the kernel $\Theta^l(\omega_{max}-\omega)$ is defined in terms of a step-function which, by definition, is not a smooth function, one cannot perform the substitution straightforwardly.
This problem can be solved introducing a smeared version of the step function $\theta_{\sigma}$ which in turns defines a smeared integration kernel $\Theta^l_{\sigma}(\omega_{max}-\omega)$ that is sufficiently smooth such that it can be reconstructed with a series of polynomials
\begin{align}
  \Theta^l_{\sigma}(\omega_{max}-\omega)
%   &
  = (\omega_{max}-\omega)^l\theta_{\sigma}(\omega_{max}-\omega)
%   \\
%   &
  = m_{B_s}^l\sum_{\tau}^{\infty}
  g_{\tau}(\omega_{max};\sigma)e^{-a\omega\tau}\;.
\end{align}
This final step allows us to write eq.~(\ref{eq:Zfunction}) in terms of products of the Euclidean correlator and the coefficients of the smeared kernel:
\begin{align}
  Z_{\sigma}^{(l)}(\boldsymbol{q}^2) = \sum_{\tau}^{\infty}
  g_{\tau}(\omega_{max};\sigma) G^{(l)}(a\tau;\boldsymbol{q})\;.
  \label{eq:Z_function_G}
\end{align}

It is important to notice that eq.~(\ref{eq:Z_function_G}) is not the same quantity showed in eq.~(\ref{eq:Zfunction}) because it is defined in terms of a smeared kernel and therefore cannot be used to compute the differential decay rate.
Moreover, hadronic spectral densities, and therefore also the structure functions $Z^{(l)}(\omega,\boldsymbol{q}^2)$, contained in the definition of the $G^{(l)}$ correlators are computed in a finite volume and as a consequence have a discrete energy spectrum, and therefore cannot be directly associated with physical observables.
\subsection{The HLT algorithm}
In the above section we discussed how one of the key ingredients for the calculation of eq.~(\ref{eq:Zfunction}) are the smeared kernel coefficients $g_{\tau}$. In this section we briefly review the method proposed in ref.~\cite{Hansen:2019idp} and we direct the reader interested in the Chebyshev method to ref.~\cite{Bailas:2020qmv,Hashimoto:2021hqu}.

\begin{figure}[tbp]
  \centering
  \includegraphics[width=\textwidth]{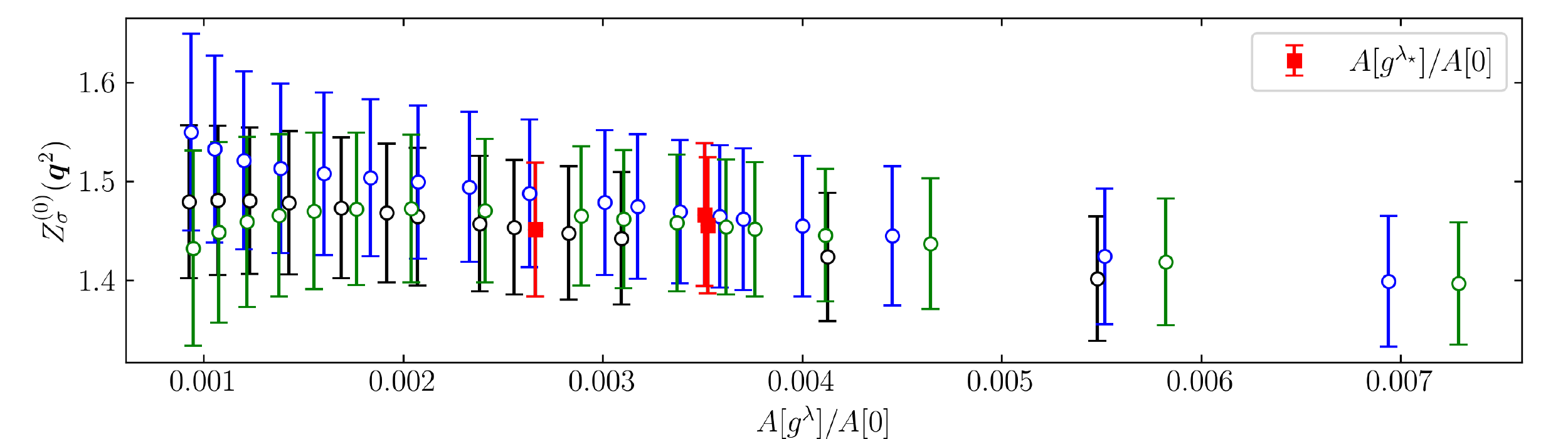}
\caption{
  Integral $\bar Z^{(0)}_\sigma(\boldsymbol{q})$ of the hadronic correlator with three kernels, plotted as a function of $A[g^\lambda]/A[0]$.
  No significant difference is observed within the statistical errors for values $A[g^{\lambda}]/A[0]$ smaller than $A[g^{\lambda_{\star}}]/A[0]$.}
  \label{fig:lambda}
\end{figure}

In the method of ref.~\cite{Hansen:2019idp}, the coefficients $g_{\tau}$ are obtained by minimising the functional
\begin{align}
  W_{\lambda}[g] = (1-\lambda)\frac{A[g]}{A[0]}+\lambda B[g]
\end{align}
where the $A[g]$ and $B[g]$ functionals are given by
\begin{align}
A[g] =
a\int_{E_0}^\infty d\omega \left\{ \Theta^l(\omega_{max}-\omega) - \sum_{\tau=1}^{\tau_{\rm max}} g_\tau \, e^{-a\omega\tau}\right\}^2\,,
\qquad
B[g] = \sum_{\tau,\tau^\prime=1}^{\tau_{\rm max}} g_{\tau} g_{\tau^\prime} \, 
\frac{{\rm Cov} \left[G(a\tau), G(a\tau^\prime) \right]}{\left[G(0)\right]^2} \,.
\label{eq:AandB}
\end{align}
The parameter $\lambda$ is a trade-off parameter which regulates the minimisation procedure between the $A[g]$ and $B[g]$ functionals.
The functional $A[g]$ is the reconstruction bias which, in the absence of errors, after the minimisation procedure gives the best polynomial approximation of the kernel $\Theta^l(\omega_{max}-\omega)$. However, considering that lattice data always have an associated statistical error, what happens is that minimising the $A[g]$ functional for small $\sigma$ results in huge values of the coefficients $g_{\tau}$.
This is an effect related to the fact that we are dealing with an ill-posed numerical problem.
The introduction of the $B[g]$ functional together with the parameter $\lambda$ solves this problem by extending the minimisation procedure to the statistical variance at the expense of a perfect kernel reconstruction.
The coefficients $g_{\tau}^{\lambda}$ that minimise $W[g]$ will then be a particular balance between the systematic and statistical errors, influenced by the $\lambda$ parameter.
Following the procedure of ref.~\cite{Hansen:2019idp}, the optimal value of $\lambda$ is obtained by finding the maximum of the functional $W[g]$ as a function of $\lambda$, finding $\lambda_{\star}$.
One can appreciate the role of $\lambda$ in balancing the statistical and systematic errors in figure~\ref{fig:lambda}.

\section{Lattice results and comparison with the OPE}
  
\begin{figure}[tbp]
  \centering
  \includegraphics[width=0.65\textwidth]{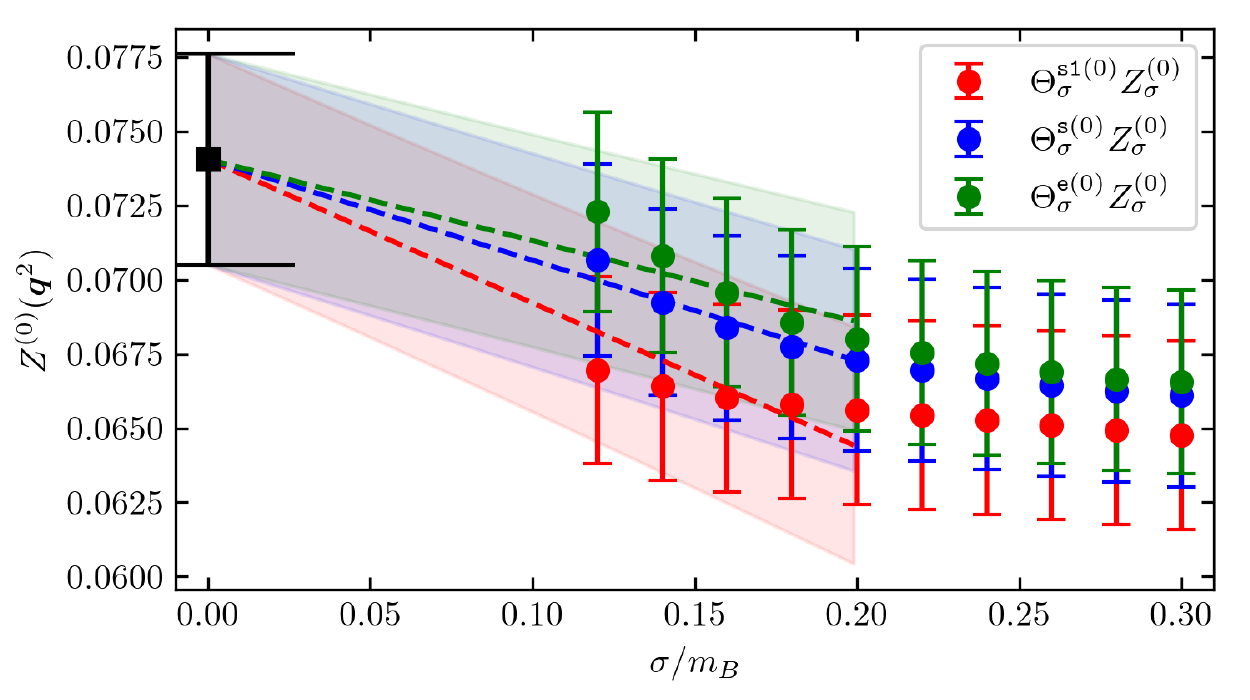}
  \caption{Combined $\sigma \rightarrow 0$ extrapolation of $Z^0_{\sigma}$ employing $10$ values of $\sigma \in [0.12 m_{B_s},0.3 m_{B_s}]$ and using the smallest $5$ to perform the fit.}
    \label{fig:extrapolation}
\end{figure}

Having found the values of the $g_{\tau}$ coefficients, either with the method of ref.~\cite{Bailas:2020qmv} or that of ref.~\cite{Hansen:2019idp}, one is then able to compute eq.~(\ref{eq:Z_function_G}). As discussed in section \ref{sec:kernel}, the quantity calculated in eq.~(\ref{eq:Z_function_G}) cannot be used to compute any physical observable.
In order to make a connection between $Z^{(l)}_{\sigma}$ and the corresponding physical quantity one has to \textit{first} perform the infinite-volume limit and only then take the $\sigma \rightarrow 0$ limit:
\begin{align}
  Z^{(l)}(\boldsymbol{q}^2)
  =  \lim_{\sigma \rightarrow 0}\left(\lim_{V \rightarrow \infty}\right)m_{B_s}^l\sum_{\tau}^{\infty}g_{\tau}(\omega,\sigma)G^{(l)}(a\tau;\boldsymbol{q})\;.
\end{align}
This is because the study of the infinite-volume limit of smeared spectral densities is well-defined only for fixed smearing functions and therefore the two limits do not commute.
However, due to the  exploratory nature of this work, we restricted our analysis only to one physical volume for each configuration ensemble, so that a $V \rightarrow \infty$ extrapolation was not possible.  This choice can be justified by the fact that our present statistical uncertainties are likely larger than finite-volume effects; we plan to investigate these effects more thoroughly in future work with simulations in multiple volumes. Figure~\ref{fig:extrapolation} shows the $\sigma \rightarrow 0$ limit taken with three smeared kernels.

After taking the $\sigma \rightarrow 0$ limit, the  $Z^{(l)}(\boldsymbol{q})$ can be used to compute the differential decay rate as shown in eq.~(\ref{eq:IncDecay}). The lattice results obtained using the JLQCD and ETM ensembles cannot be compared with each other since they use different quark masses in their simulations. However, both of them can be compared with the analytic predictions from the OPE.
\begin{figure}[h]
  \centering
  \includegraphics[width=10cm]{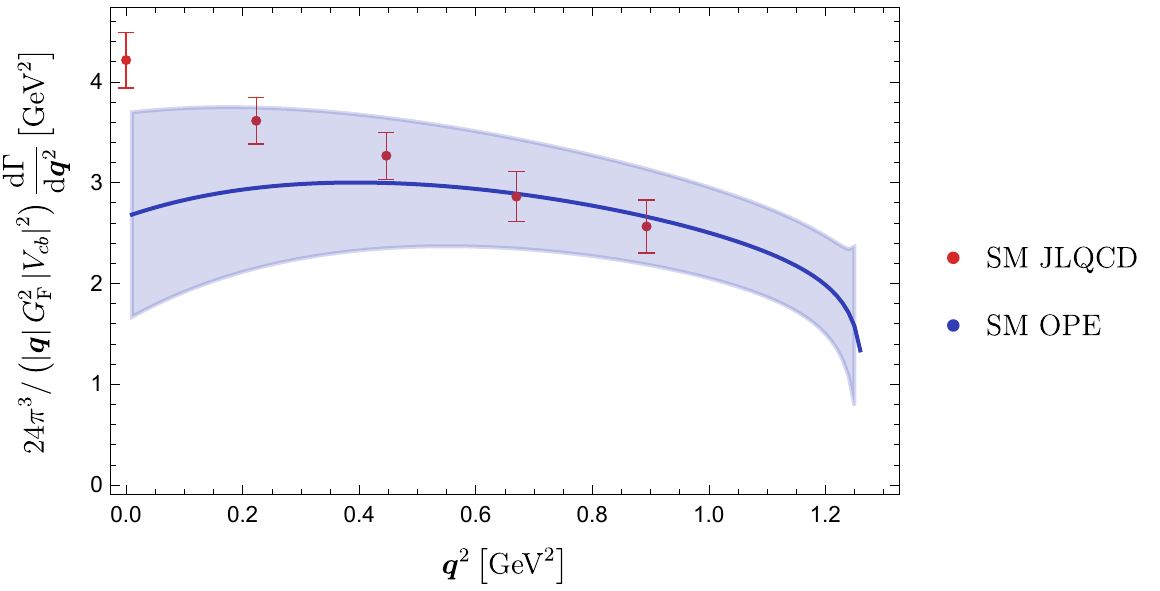}
  \includegraphics[width=10cm]{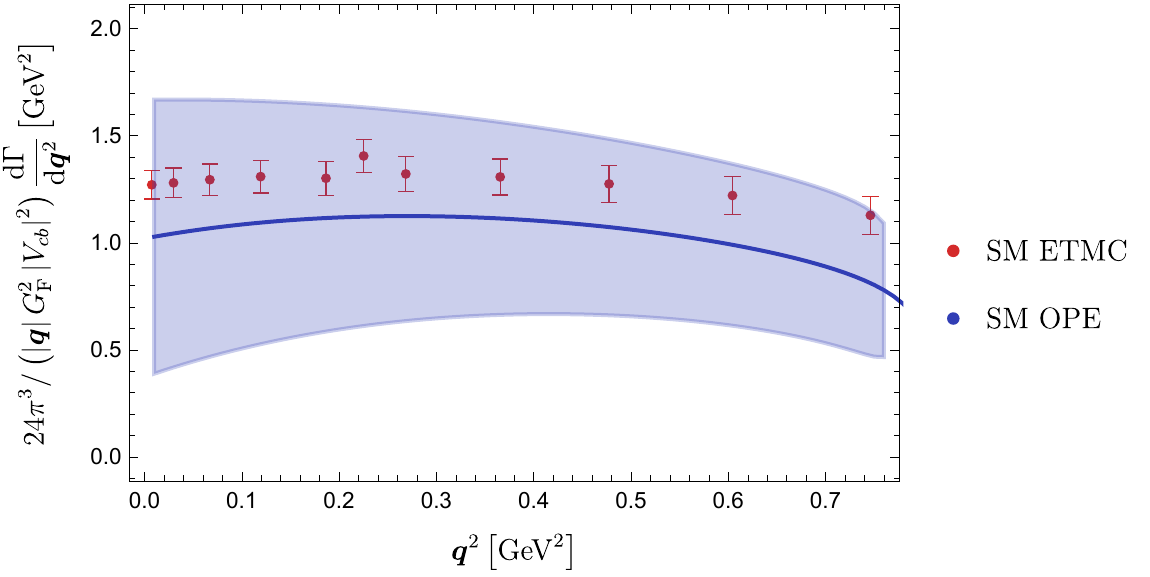}  
\caption{Differential $\boldsymbol{q}^2$ spectrum, divided by $|\boldsymbol{q}|$, in the SM. Comparison of OPE predictions with lattice QCD results from the JLQCD (top panel) and ETM (bottom panel) ensembles are shown.}
\label{fig:Comp}
\end{figure}
As shown in figure~\ref{fig:Comp}, the agreement between the OPE curve (in blue) and the lattice data (in red) is very good, for both the JLQCD and the ETM ensembles.

One can then obtain the inclusive decay rate by performing the integration over $\boldsymbol{q}$. In table~\ref{tab:results} we show the final results obtained using the JLQCD and ETM ensembles compared with the OPE predictions. We also show the results of the first lepton energy moment $\langle E_l \rangle$ and the first hadronic mass moment $\langle M_X^2 \rangle$.
\begin{table}[tbp]
  \begin{center}
  \begin{tabular}{c|ll}\hline
 &JLQCD & OPE%(\small JLQCD) 
   \\ \hline
 $\Gamma/|V_{cb}^2|  \times 10^{13}$ (GeV)  &  $4.46(21)$ &  5.7(9)  \\
 $\langle E_\ell \rangle$ (GeV) & 0.650(40)&  0.626(36) \\
%  $\langle E_\ell^2 \rangle$ (GeV) & &0.431(57)  \\
% $\langle E_\ell^2 \rangle-\langle E_\ell \rangle^2$(GeV$^2$) & & 0.044(11)\\
 $\langle M_X^2\rangle$ (GeV$^2$)&  3.75(31)& 4.22(30)  \\
   \hline
  \end{tabular} 
  \quad
  \begin{tabular}{ll}\hline
    ETMC & OPE %(\small ETMC) 
    \\ \hline
    0.987(60) & 1.20(46) \\
    0.491(15)& 0.441(43) \\
    3.62(14)&  4.32(56)\\
    \hline
  \end{tabular}
  \end{center}
  \caption{Total widths and moments.}
  \label{tab:results}
\end{table}
What can be concluded from this comparison is that the lattice data show a good agreement with the OPE curves, especially at low $\boldsymbol{q}$ and the final results are compatible within the uncertainties.

These findings provide an important and non-trivial test of the method used in this work, making us optimistic about the possibility of soon having a full lattice QCD computation of inclusive semileptonic decays, at a level of precision competitive with those from the OPE. This will certainly give a big contribution towards a more precise inclusive determination of $|V_{cb}|$, which will hopefully clarify the origin of the current $3 \sigma$ tension between the inclusive and exclusive determination of this quantity.
Another reason why a full inclusive lattice QCD computation is important is the possibility of testing the quark-hadron duality on which the OPE is based, by comparing calculations of the same physical observable with both methods.

\section{Conclusions}
We presented our recent study of the inclusive semileptonic decay of the $B_s$ meson from lattice QCD~\cite{Gambino:2022dvu}, which is based on the strategy proposed in ref.~\cite{Gambino:2020crt}. As we discussed, our calculation is based on the numerical reconstruction of the integration kernel that regulates the integral over phase space of the structure functions contributing to the hadronic tensor. We computed four-point Euclidean correlation functions which were obtained from two different gauge ensembles, based on different discretisation schemes for both gauge and quark fields and at $b$-quark masses lighter than in nature. We used two different methods for the kernel reconstructions, which were applied to the correlation functions in order to obtain an inclusive calculation of the semileptonic decay rate.
The results that we obtained are consistent and can be successfully compared with analytic predictions from the OPE approach. This opens the path for systematic \emph{ab initio} non-perturbative studies of inclusive decays on the lattice, which hopefully will help solve the $V_{cb}$ puzzle.

\acknowledgments
The numerical calculations of the JLQCD collaboration were performed on the SX-Aurora TSUBASA at the High Energy Accelerator Research Organization (KEK) under its Particle, Nuclear and Astrophysics Simulation Program, as well as on the Oakforest-PACS supercomputer operated by the Joint Center for Advanced High Performance Computing (JCAHPC). We thank the members of the JLQCD collaboration for sharing the computational framework and lattice data, and Takashi~Kaneko in particular for providing the numerical data for the exclusive decay form factors. The numerical simulations of the ETM collaboration were run on machines of the Consorzio Interuniversitario per il Calcolo Automatico dell'Italia Nord Orientale (CINECA) under the specific initiative INFN-LQCD123.
The work of S.H. is supported in part by JSPS KAKENHI Grant Number 22H00138 and by the Post-K and Fugaku supercomputer project through the Joint Institute for Computational Fundamental Science (JICFuS). The work of P.G., S.M., F.S., S.S. is supported by the Italian Ministry of Research (MIUR) under grant PRIN 20172LNEEZ. This project has received funding from the Swiss National Science Foundation (SNF) under contract 200020\_204428.
%\item Plenary review talk: $20\pm5$ pages
%\item Plenary topical talk: $15\pm3$ pages
%\item Parallel talk or poster: $7\pm2$ pages

\bibliographystyle{JHEP}
\bibliography{template_lat22}

\providecommand{\href}[2]{#2}\begingroup\raggedright\begin{thebibliography}{10}

\bibitem{ParticleDataGroup:2020ssz}
{\scshape Particle Data Group} collaboration, \emph{{Review of Particle
  Physics}}, \href{https://doi.org/10.1093/ptep/ptaa104}{\emph{PTEP} {\bfseries
  2020} (2020) 083C01}.

\bibitem{HFLAV:2019otj}
{\scshape HFLAV} collaboration, \emph{{Averages of $b$-hadron, $c$-hadron, and
  $\tau $-lepton properties as of 2018}},
  \href{https://doi.org/10.1140/epjc/s10052-020-8156-7}{\emph{Eur. Phys. J. C}
  {\bfseries 81} (2021) 226}
  [\href{https://arxiv.org/abs/1909.12524}{{\ttfamily 1909.12524}}].

\bibitem{Gambino:2019sif}
P.~Gambino, M.~Jung and S.~Schacht, \emph{{The $V_{cb}$ puzzle: An update}},
  \href{https://doi.org/10.1016/j.physletb.2019.06.039}{\emph{Phys. Lett. B}
  {\bfseries 795} (2019) 386}
  [\href{https://arxiv.org/abs/1905.08209}{{\ttfamily 1905.08209}}].

\bibitem{Gambino:2020jvv}
P.~Gambino et~al., \emph{{Challenges in semileptonic {$B$} decays}},
  \href{https://doi.org/10.1140/epjc/s10052-020-08490-x}{\emph{Eur. Phys. J. C}
  {\bfseries 80} (2020) 966}
  [\href{https://arxiv.org/abs/2006.07287}{{\ttfamily 2006.07287}}].

\bibitem{Wilson:1969zs}
K.G.~Wilson, \emph{{Nonlagrangian models of current algebra}},
  \href{https://doi.org/10.1103/PhysRev.179.1499}{\emph{Phys. Rev.} {\bfseries
  179} (1969) 1499}.

\bibitem{Kadanoff:1969zz}
L.P.~Kadanoff, \emph{{Operator Algebra and the Determination of Critical
  Indices}}, \href{https://doi.org/10.1103/PhysRevLett.23.1430}{\emph{Phys.
  Rev. Lett.} {\bfseries 23} (1969) 1430}.

\bibitem{FlavourLatticeAveragingGroup:2019iem}
{\scshape Flavour Lattice Averaging Group} collaboration, \emph{{FLAG Review
  2019: Flavour Lattice Averaging Group (FLAG)}},
  \href{https://doi.org/10.1140/epjc/s10052-019-7354-7}{\emph{Eur. Phys. J. C}
  {\bfseries 80} (2020) 113}
  [\href{https://arxiv.org/abs/1902.08191}{{\ttfamily 1902.08191}}].

\bibitem{Gambino:2022dvu}
P.~Gambino, S.~Hashimoto, S.~M\"achler, M.~Panero, F.~Sanfilippo, S.~Simula
  et~al., \emph{{Lattice QCD study of inclusive semileptonic decays of heavy
  mesons}}, \href{https://doi.org/10.1007/JHEP07(2022)083}{\emph{JHEP}
  {\bfseries 07} (2022) 083}
  [\href{https://arxiv.org/abs/2203.11762}{{\ttfamily 2203.11762}}].

\bibitem{Hashimoto:2017wqo}
S.~Hashimoto, \emph{{Inclusive semi-leptonic $B$ meson decay structure
  functions from lattice QCD}},
  \href{https://doi.org/10.1093/ptep/ptx052}{\emph{PTEP} {\bfseries 2017}
  (2017) 053B03} [\href{https://arxiv.org/abs/1703.01881}{{\ttfamily
  1703.01881}}].

\bibitem{Gambino:2020crt}
P.~Gambino and S.~Hashimoto, \emph{{Inclusive Semileptonic Decays from Lattice
  QCD}}, \href{https://doi.org/10.1103/PhysRevLett.125.032001}{\emph{Phys. Rev.
  Lett.} {\bfseries 125} (2020) 032001}
  [\href{https://arxiv.org/abs/2005.13730}{{\ttfamily 2005.13730}}].

\bibitem{Nakayama:2017lav}
{\scshape JLQCD} collaboration, \emph{{Determination of charm quark mass from
  temporal moments of charmonium correlator with M\"obius domain-wall
  fermion}}, \href{https://doi.org/10.22323/1.256.0192}{\emph{PoS} {\bfseries
  LATTICE2016} (2017) 192} [\href{https://arxiv.org/abs/1702.01498}{{\ttfamily
  1702.01498}}].

\bibitem{Colquhoun:2022atw}
{\scshape JLQCD} collaboration, \emph{{Form factors of {$B\to\pi\ell\nu$} and a
  determination of {$|V_{ub}|$} with M{\"o}bius domain-wall-fermions}},
  \href{https://doi.org/10.1103/PhysRevD.106.054502}{\emph{Phys. Rev. D}
  {\bfseries 106} (2022) 054502}
  [\href{https://arxiv.org/abs/2203.04938}{{\ttfamily 2203.04938}}].

\bibitem{Baron:2010bv}
R.~Baron et~al., \emph{{Light hadrons from lattice QCD with light ($u$,$d$),
  strange and charm dynamical quarks}},
  \href{https://doi.org/10.1007/JHEP06(2010)111}{\emph{JHEP} {\bfseries 06}
  (2010) 111} [\href{https://arxiv.org/abs/1004.5284}{{\ttfamily 1004.5284}}].

\bibitem{ETM:2010cqp}
{\scshape ETM} collaboration, \emph{{Light hadrons from $N_f=2+1+1$ dynamical
  twisted mass fermions}},
  \href{https://doi.org/10.22323/1.105.0123}{\emph{PoS} {\bfseries LATTICE2010}
  (2010) 123} [\href{https://arxiv.org/abs/1101.0518}{{\ttfamily 1101.0518}}].

\bibitem{Frezzotti:2000nk}
{\scshape Alpha} collaboration, \emph{{Lattice QCD with a chirally twisted mass
  term}}, \href{https://doi.org/10.1088/1126-6708/2001/08/058}{\emph{JHEP}
  {\bfseries 08} (2001) 058}
  [\href{https://arxiv.org/abs/hep-lat/0101001}{{\ttfamily hep-lat/0101001}}].

\bibitem{Frezzotti:2003xj}
R.~Frezzotti and G.C.~Rossi, \emph{{Twisted mass lattice QCD with mass
  nondegenerate quarks}},
  \href{https://doi.org/10.1016/S0920-5632(03)02477-0}{\emph{Nucl. Phys. B
  Proc. Suppl.} {\bfseries 128} (2004) 193}
  [\href{https://arxiv.org/abs/hep-lat/0311008}{{\ttfamily hep-lat/0311008}}].

\bibitem{Frezzotti:2003ni}
R.~Frezzotti and G.C.~Rossi, \emph{{Chirally improving Wilson fermions. 1.
  $O(a)$ improvement}},
  \href{https://doi.org/10.1088/1126-6708/2004/08/007}{\emph{JHEP} {\bfseries
  08} (2004) 007} [\href{https://arxiv.org/abs/hep-lat/0306014}{{\ttfamily
  hep-lat/0306014}}].

\bibitem{EuropeanTwistedMass:2014osg}
{\scshape ETM} collaboration, \emph{{Up, down, strange and charm quark masses
  with $N_f = 2+1+1$ twisted mass lattice QCD}},
  \href{https://doi.org/10.1016/j.nuclphysb.2014.07.025}{\emph{Nucl. Phys. B}
  {\bfseries 887} (2014) 19} [\href{https://arxiv.org/abs/1403.4504}{{\ttfamily
  1403.4504}}].

\bibitem{Bailas:2020qmv}
G.~Bailas, S.~Hashimoto and T.~Ishikawa, \emph{{Reconstruction of smeared
  spectral function from Euclidean correlation functions}},
  \href{https://doi.org/10.1093/ptep/ptaa044}{\emph{PTEP} {\bfseries 2020}
  (2020) 043B07} [\href{https://arxiv.org/abs/2001.11779}{{\ttfamily
  2001.11779}}].

\bibitem{Hansen:2019idp}
M.~Hansen, A.~Lupo and N.~Tantalo, \emph{{Extraction of spectral densities from
  lattice correlators}},
  \href{https://doi.org/10.1103/PhysRevD.99.094508}{\emph{Phys. Rev. D}
  {\bfseries 99} (2019) 094508}
  [\href{https://arxiv.org/abs/1903.06476}{{\ttfamily 1903.06476}}].

\bibitem{Hashimoto:2021hqu}
S.~Hashimoto and G.~Bailas, \emph{{Composition of the inclusive semi-leptonic
  decay of $B$ meson}}, \href{https://doi.org/10.22323/1.396.0534}{\emph{PoS}
  {\bfseries LATTICE2021} (2022) 534}
  [\href{https://arxiv.org/abs/2112.09295}{{\ttfamily 2112.09295}}].

\end{thebibliography}\endgroup
\end{document}